\documentclass[11pt,twoside]{article}
\usepackage{vdbbook}
\usepackage{amssymb}

\resetcounters

\begin{document}


\title{Physical Conditions in the Interstellar Medium of High-Redshift Galaxies:\\ Mass Budget and Gas Excitation}
\author{Dominik A.\ Riechers$^1$
\affil{$^1$Cornell University, 220 Space Sciences Building, Ithaca, NY, USA}}


\begin{abstract}

Following the first pioneering efforts in the 1990s that have focused
on the detection of the molecular interstellar medium in high redshift
galaxies, recent years have brought great advances in our
understanding of the actual physical properties of the gas that set
the conditions for star formation. Observations of the ground-state CO
$J$=1$\to$0 line have furnished crucial information on the total
masses of the gas reservoirs, as well as reliable dynamical mass and
size estimates. Detailed studies of rotational ladders of CO have
provided insight on the temperature and density of the
gas. Investigations of the very dense gas associated with actively
star-forming regions in the interstellar medium, most prominently
through HCN and HCO$^+$, have enabled a better understanding of the
nature of the extreme starbursts found in many high-redshift galaxies,
which exceed the star formation rates of their most active present-day
counterparts by an order of magnitude.  Key progress in this area has
been made through targeted studies of few, well-selected systems with
current facilities. With the completion of the Karl G.\ Jansky Very
Large Array and the Atacama Large (sub)Millimeter Array, it will
become possible to develop a more general framework for the
interpretation of these investigations based on unbiased studies of
``normal'' star-forming galaxies back to the earliest cosmic epochs.
\end{abstract}

\section{Introduction}
\label{intro}

The past two decades have shown tremendous progress in studies of
galaxy evolution out to high redshift. We have now understood that
galaxies formed a lot more stars in the past than they do today (by an
order of magnitude or more by $z$$\simeq$1; e.g., Madau et al.\ 1996;
Lilly et al.\ 1996), and that the buildup of stellar mass in galaxies
through cosmic times took place through both brief, but very intense
starburst periods lasting tens to 100\,Myr and steady, more quiescent
rates of star formation lasting typically half a Gyr or longer (e.g.,
Daddi et al.\ 2010a, Genzel et al.\ 2010). In many cases, the most
infrared-luminous starbursts are driven by major mergers. The most
extreme starbursts observed at earlier cosmic times are typically by
an order of magnitude more infrared-luminous than the most extreme
examples observed today (e.g., Blain et al.\ 2002). In addition, the
luminosity threshold above which star-forming galaxies are dominantly
major mergers has likely been higher by an order of magnitude or more
in the past as well (e.g., Daddi et al.\ 2007). In recent years, it
has been found that this increased activity in galaxies at high
redshift can be understood in the context of a higher gas content of
galaxies on average, and that even fairly typical high-redshift
galaxies contain high fractions of molecular gas (e.g., Daddi et
al.\ 2008, 2010b; Tacconi et al.\ 2010). Thus, studies of molecular
gas in distant galaxies have become an important means towards
understanding cosmic star formation at early epochs.

Molecular gas is an important probe of the physical conditions in
distant galaxies. Great progress has been made since the initial
detections of CO at high redshift more than two decades ago (Brown \&
Vanden Bout 1991; Solomon et al.\ 1992). Significant samples of
different high-redshift galaxy populations have been detected in CO
emission, and detailed follow-up campaigns of subsamples at spatial
resolutions of up to 1\,kpc have provided detailed insight on the
spatial distribution, mass density, and dynamical structure of the
gas. Studies of the gas excitation have constrained the physical
properties of the interstellar medium, and the detection of other
molecules such as HCN, HCO$^+$, HNC, CN, and H$_2$O has enabled
studies of the chemical composition (see Solomon \& Vanden Bout 2005;
Omont 2007; Carilli \& Walter 2013 for comprehensive reviews of the
subject). The future of the field is bright, now that we have entered
a new era in such investigations with the advent of powerful new
facilities such as the Karl G.\ Jansky Very Large Array (JVLA) and the
Atacama Large sub/Millimeter Array (ALMA). In the context of this
volume, this article highlights some key aspects related to the
determination of total molecular gas masses in high-$z$ galaxies, and
discusses what physical properties can be probed through studies of
the gas composition and excitation.

\section{Detections of Molecular Gas at High Redshift:\ a Brief Summary}
\label{intro}

\begin{figure}[t]
\centering
\includegraphics[scale=0.65]{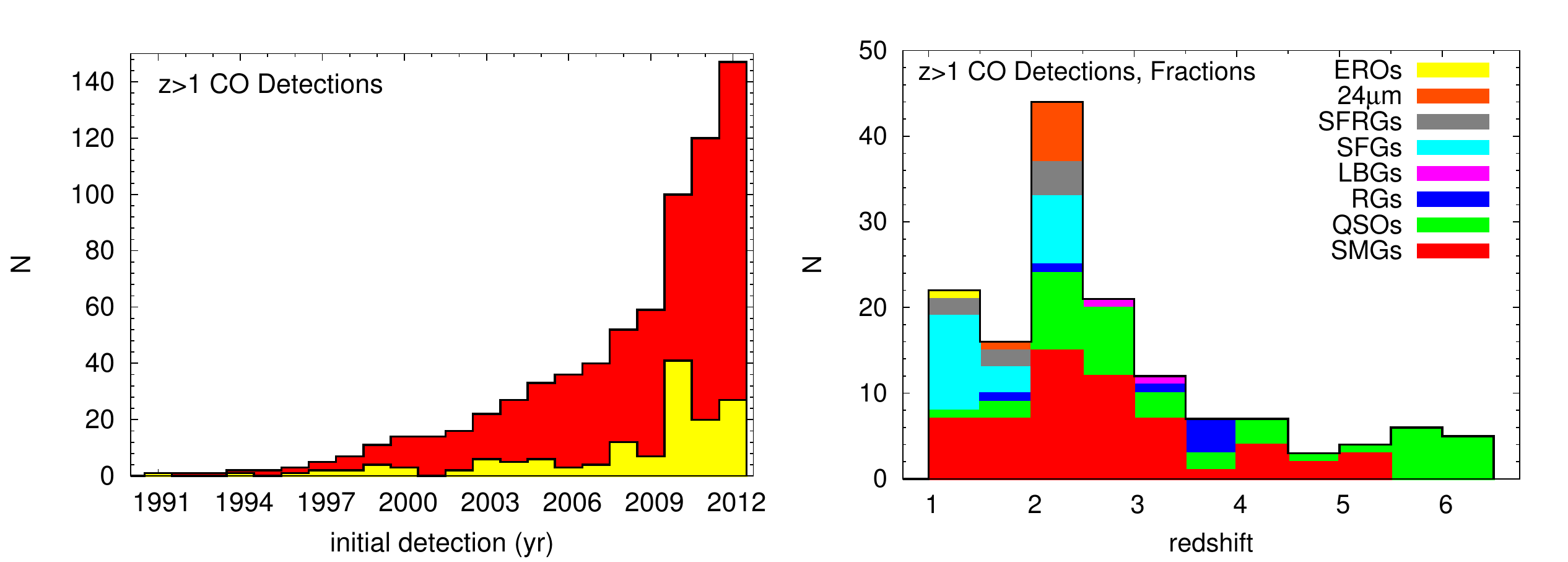} 
\caption{Detections of CO emission in $z$$>$1 galaxies as of
  2012. {\em Left:} Total number of detections (red) and detections
  per year (yellow) since the initial detection in 1991/1992. {\em
    Right:} Detections as a function of redshift, and color encoded by
  galaxy type (figure updated from Riechers 2011a, 2012).
}
\label{f1}
\end{figure}

To date,\footnote{This census is based on data published in the
  peer-reviewed literature by the end of 2012, and updates previous
  compilations shown by Riechers (2011a, 2012). It does not contain
  galaxies detected in [CII] emission but not CO (e.g., Stacey et
  al.\ 2010; Venemans et al.\ 2012).} molecular gas (most commonly CO)
has been detected in $\sim$150 galaxies at $z$$>$1 (Fig.~\ref{f1}),
back to only 870 million years after the Big Bang (corresponding to
$z$=6.42; e.g., Walter et al.\ 2003, 2004; Bertoldi et al.\ 2003;
Riechers et al.\ 2009a). Except for few sources highly magnified by
strong gravitational lensing (e.g., Baker et al.\ 2004; Coppin et
al.\ 2007; Riechers et al.\ 2010a), these are massive galaxies hosting
large molecular gas reservoirs (typically 10$^{10}$\,M$_\odot$ or
more), commonly with high gas fractions of (at least) tens of per
cent. Approximately 20\% of the detected systems are massive, gas-rich
optically/near-infrared selected star forming galaxies (SFGs; e.g.,
Daddi et al.\ 2010b; Tacconi et al.\ 2010), and 30\% each are
far-infrared-luminous, star-bursting quasars (QSOs; e.g., Wang et
al.\ 2010; Riechers et al.\ 2006a; Riechers 2011b) and submillimeter
galaxies (SMGs; e.g., Greve et al.\ 2005; Tacconi et al.\ 2008;
Riechers et al.\ 2010b; Fig.~\ref{f1}). The rest of CO-detected
high-redshift galaxies are limited samples of galaxies selected
through a variety of techniques, such as Extremely Red Objects (EROs),
Star-Forming Radio-selected Galaxies (SFRGs), 24\,$\mu$m-selected
galaxies, gravitationally lensed Lyman-break galaxies (LBGs), and
radio galaxies (RGs; see Carilli \& Walter 2013 for a recent
summary). Besides CO, the high-density gas tracers HCN, HCO$^+$, HNC,
CN, and H$_2$O were detected towards a small subsample of these
galaxies (e.g., Solomon et al.\ 2003; Vanden Bout et al.\ 2004;
Riechers et al.\ 2006b, 2007a, 2010c, 2011a; Guelin et al.\ 2007;
Omont et al.\ 2011).

\section{Total Molecular Gas Masses}
\label{mgas}

\begin{figure}[t]
\centering
\includegraphics[scale=0.22]{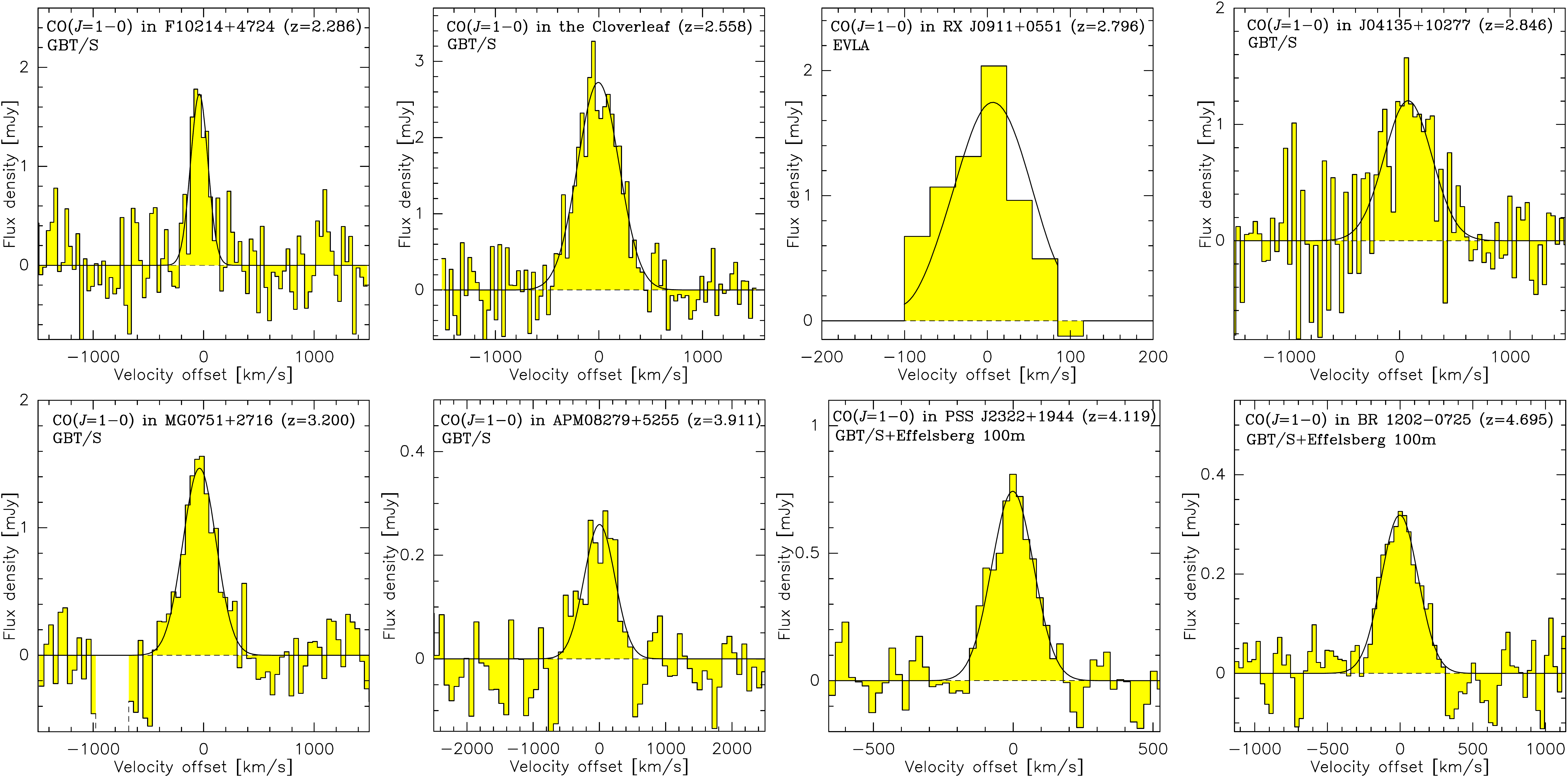} 
\caption{CO($J$=1$\to$0) spectroscopy of high-redshift quasar host
  galaxies with the Green Bank Telescope, the Effelsberg 100\,m, and
  the JVLA (Riechers et al.\ 2006a, 2011b). F10214+4724 is the galaxy
  that was initially studied by Brown \& Vanden Bout (1991) and
  Solomon et al.\ (1992) in the CO($J$=3$\to$2) line, but only
  recently, detection of CO($J$=1$\to$0) in this system became
  possible due to previous tuning restrictions. The spectral coverage
  of the CO($J$=1$\to$0) spectrum of RX\,J0911+0551 was limited by the
  bandwidth of the old VLA correlator. All galaxies shown here except
  BR\,1202--0725 are gravitationally lensed. The measured
  CO($J$=1$\to$0) luminosity is used to determine the total molecular
  gas masses of these objects.}
\label{f2}
\end{figure}

Molecular gas is the fuel for star formation, and thus, a key aspect
in studies of galaxy evolution. Measurements of total molecular gas
masses (and thereby, gas fractions) of galaxies provide the means to
understand in what phase of their evolution galaxies are, as they
place direct constraints on the amount of material that is left for
future star formation, and thus, on how long star formation can be
maintained at the current rate, and on how much stellar mass can be
assembled without a source of external gas supply.

In the nearby universe, molecular gas masses are most commonly
determined based on observations of CO($J$=1$\to$0) emission. To
determine the mass in H$_2$ gas from the CO($J$=1$\to$0) luminosity is
not trivial, as the conversion factor ($\alpha_{\rm CO}$) between both
quantities depends on the physical conditions in the gas, as well as
on the gas phase metallicity (see review by Bolatto et
al.\ 2013). Values for $\alpha_{\rm CO}$ range from
0.3--1.3\,M$_\odot$\,(K\,km\,s$^{-1}$\,pc$^2$)$^{-1}$ in the most
intense starbursts (e.g., Downes \& Solomon 1998) to
3.5--4.6\,M$_\odot$\,(K\,km\,s$^{-1}$\,pc$^2$)$^{-1}$ in quiescently
star-forming galaxies (e.g., Solomon \& Barrett 1991), and can be even
higher in low-metallicity galaxies (e.g., Leroy et
al.\ 2011). However, the CO($J$=1$\to$0) line luminosity is still the
best-calibrated estimator available in the local universe, and given
the limited direct constraints available at high redshift, the most
reliable diagnostic to be utilized in distant galaxies as well. Also,
our theoretical understanding of variations in $\alpha_{\rm CO}$ among
galaxies near and far has been improving in recent years (see, e.g.,
Narayanan 2013, this volume). Due to the redshifting of molecular
lines, high-$z$ galaxies are most commonly detected in $J$$\geq$3 CO
lines, which are shifted to the 3\,mm observing window that is
accessible with telescopes that are used for the detection of
CO($J$=1$\to$0) emission in nearby galaxies. Besides $\alpha_{\rm
  CO}$, the determination of gas masses from $J$$\geq$3 CO lines bears
the additional uncertainty of gas excitation, which needs to be known
to infer the CO($J$=1$\to$0) line luminosity based on the brightness
of higher-$J$ lines. Without a correction, the CO($J$=1$\to$0) line
luminosity, and thus, the gas mass, may be underestimated by up to a
factor of a few.

\begin{figure}[t]
\vspace{-10mm}
\centering
\includegraphics[scale=0.3]{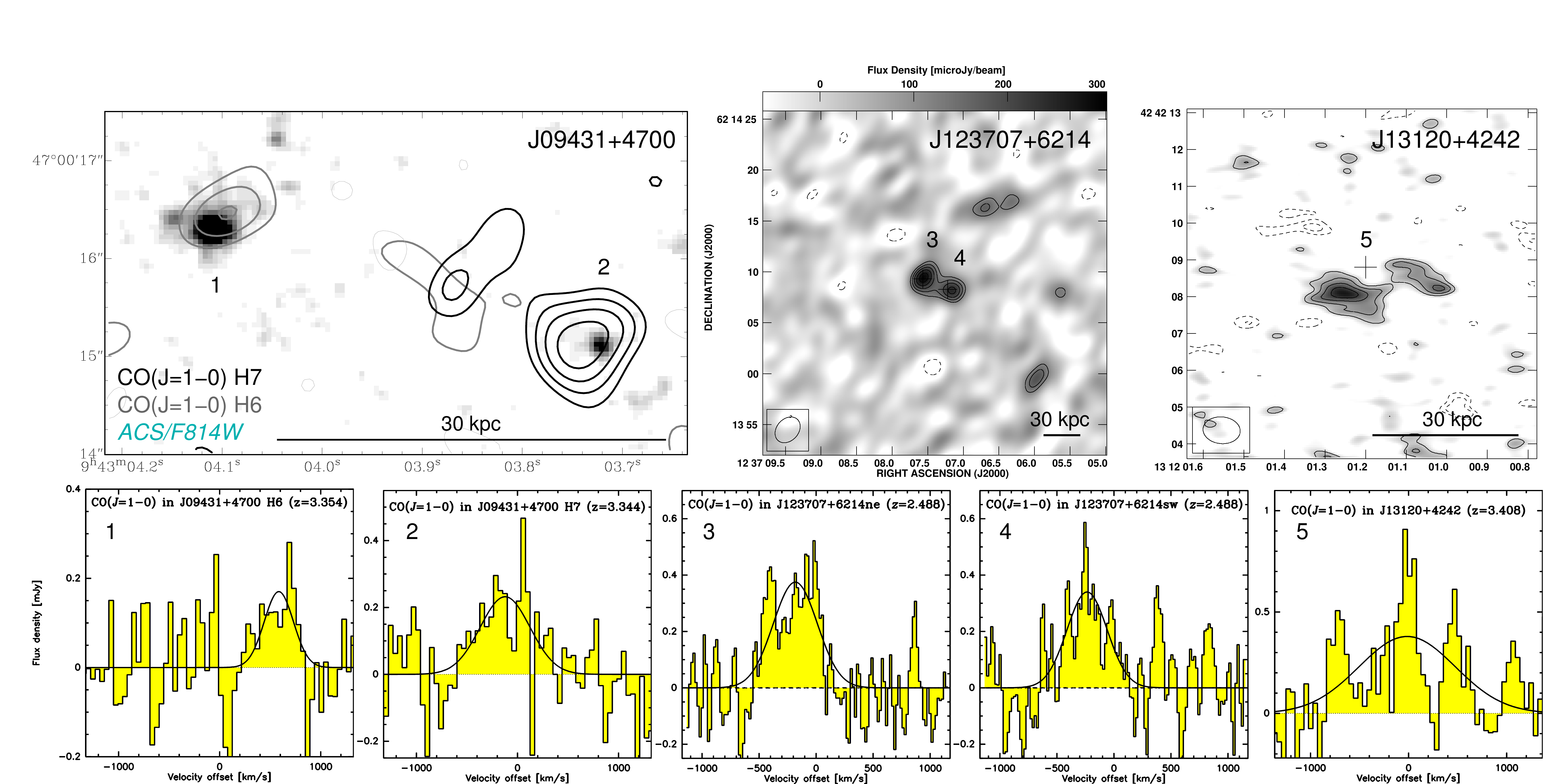} 
\caption{CO($J$=1$\to$0) observations of submillimeter galaxies along
  the ``merger sequence'' with the JVLA (Riechers et al. 2011c,
  2011d). SMGs show complex gas morphologies that can extend over
  $>$10 kpc scales, and commonly consist of multiple components. The
  gas distribution and kinematics of the majority of SMGs are as
  expected for major, gas-rich mergers. The observed diversity in
  these properties is consistent with different merger stages, as
  shown here for three examples. {\em Left:} The two gas-rich galaxies
  in this SMG are separated by tens of kpc and
  $\sim$700\,km\,s$^{-1}$, representing an early merger stage. {\em
    Middle:} This SMG still shows two separated components, but at
  similar velocity, representing a more advanced merger stage. {\em
    Right:} This SMG shows a single, complex extended gas structure
  with multiple velocity components, representing a fairly late merger
  stage.}
\label{f3}
\end{figure}

To overcome this limitation, direct studies of CO($J$=1$\to$0)
emission in high-redshift galaxies, redshifted down to short radio
wavelengths (6\,mm--2\,cm) accessible with the Green Bank Telescope
(GBT) and the JVLA have become more common in recent years. After
initial detections in less than a handful of galaxies (e.g., Carilli
et al.\ 2002; Riechers et al.\ 2006a; Hainline et al.\ 2006), the
increased frequency coverage and spectroscopic capabilities of the
JVLA upgrade, and the availability of the Zpectrometer wide-band
instrument with stable spectral baselines at the GBT have yielded tens
of detections in galaxies at redshifts of $z$$\gtrsim$1.5,
encompassing quasar host galaxies,\footnote{Based on higher spatial
  resolution CO($J$=3$\to$2) observations than previously available,
  the CO emission in J4135+10277 shown in Fig.~\ref{f2} has recently
  been found to not be directly associated with the quasar host, but
  with a nearby, optically faint dust-obscured starburst at the same
  redshift. This gas-rich star-forming galaxy will likely merge with
  the quasar host galaxy in the future (Riechers 2013; see Hainline et
  al.\ 2004 for original CO $J$=3$\to$2 detection).} SMGs, LBGs, SFGs,
and RGs (e.g., Figs.~\ref{f2}, \ref{f3}; Riechers et al.\ 2010a,
2011b, 2011c, 2011d, 2011e; Ivison et al.\ 2010, 2011, 2012; Harris et
al.\ 2010, 2012; Frayer et al.\ 2011; Aravena et al.\ 2010). These
studies have revealed a wide range in gas excitation properties,
showing that the gas masses derived based on the brightness of
higher-$J$ CO lines were biased low by factors of two or more for many
of the systems (e.g., Riechers et al.\ 2011c, 2011d; Ivison et
al.\ 2011), while others are highly excited, and thus, did not require
significant corrections (typically quasar hosts; e.g., Riechers et
al.\ 2006a, 2009b, 2011b; see next section). Spatially resolved
mapping with the JVLA also shows that submillimeter-selected galaxies
commonly show gas reservoirs that are significantly more spatially
extended in CO($J$=1$\to$0) than in higher-$J$ lines, indicating the
presence of significant amounts of cold gas with low excitation (e.g.,
Riechers et al.\ 2011c; Ivison et al.\ 2011). This potentially
requires a different conversion factor $\alpha_{\rm CO}$ for different
components of the gas, which would imply an even larger correction to
gas masses derived from $J$$\geq$3 CO lines alone. It also implies
that accurate sizes of the gas reservoirs and gas dynamical masses for
these galaxies can only be derived based on CO($J$=1$\to$0) imaging.
In contrast, quasar host galaxies typically do not show evidence for
significant fractions of spatially extended, low-excitation gas, which
may suggest that gas masses derived from $J$$\geq$3 CO lines require
little corrections (Riechers et al.\ 2006a, 2011b). Studies of more
quasar host galaxies at high spatial resolution in both low- and
high-$J$ CO transitions are desirable to investigate this apparent
difference between quasars and SMGs with comparable far-infrared
luminosities in more detail, and to better understand the role of both
populations in the context of the evolution of massive galaxies. The
JVLA and ALMA are the ideal tools for such investigations.

\section{Molecular Gas Excitation}
\label{coex}

\subsection{CO}

\begin{figure}[t]
\centering
\includegraphics[scale=0.44]{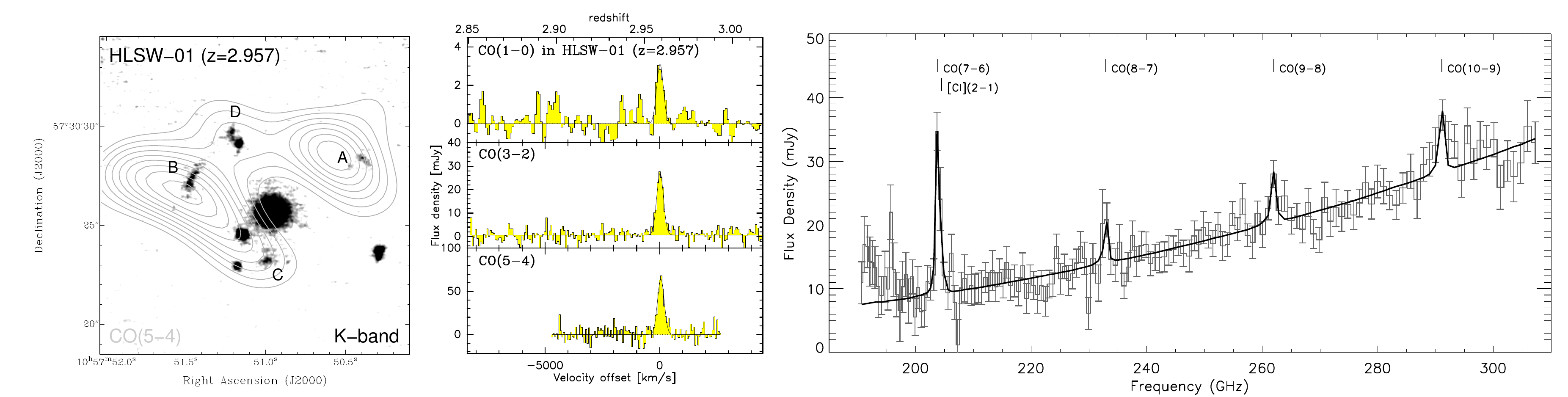} 
\caption{Multi-line CO spectroscopy of the strongly lensed
  Herschel-selected SMG HLSW--01 ($z$=2.957; Riechers et al.\ 2011e;
  Scott et al.\ 2011). {\em Left:} CO($J$=5$\to$4) emission contours
  on a 2.2\,$\mu$m continuum image, showing the lensed images of the
  SMG (labeled A to D) and the group of $z$$\simeq$0.6 galaxies
  responsible for the lensing magnification. {\em Middle:} CO
  $J$=1$\to$0, 3$\to$2, and 5$\to$4 emission lines. {\em Right:} CO
  $J$=7$\to$6 to 10$\to$9 emission lines. CO($J$=7$\to$6) is partially
  blended with the upper [CI] fine structure line. The relative
  strength of the different CO lines, covering the excitation ladder
  from $J$=1 to 10, provide detailed constraints on the excitation of
  the molecular gas.}
\label{f4}
\end{figure}

To understand the excitation properties of the molecular gas
reservoirs of high-redshift galaxies in more detail, it is necessary
to observe multiple molecular lines, most commonly by covering three
or more transitions of the rotational ladder of CO. In most cases,
these studies are currently limited to the study of spatially
integrated properties, rather than variations between different
regions of galaxies as done in the most detailed studies of nearby
galaxies (e.g., Wei\ss\ et al.\ 2005).  In the four best-studied
high-redshift systems, all of which are strongly gravitationally
lensed, seven or more CO lines have been detected (e.g.,
Fig.~\ref{f4}; Riechers et al.\ 2006a, 2011b, 2011e; Wei\ss\ et
al.\ 2007; Bradford et al.\ 2009, 2011; Scott et al.\ 2011; Danielson
et al.\ 2011). By modeling the intensity of different rotational
levels of CO relative to equilibrium (in which case all lines would
have the same brightness temperature, and the line fluxes would scale
with $\nu_{\rm line}^2$), it is possible to constrain the physical
properties of the gas (in particular its density and kinetic
temperature). The CO lines, when observed in emission, can be excited
either by collisions with other molecules, or by the ambient radiation
field (e.g., Meijerink \& Spaans 2005). In lack of detailed
constraints on the gas distribution and local radiation field, the
most common approach is to model the collisional excitation of CO
using the large velocity gradient (LVG) approximation (e.g., Scoville
\& Solomon 1974; Goldreich \& Kwan 1974). This method utilizes an
escape probability formalism (i.e., photons produced locally can only
be absorbed locally) resulting from a strong velocity gradient, which
helps to minimize the number of free parameters in models of the
collisional line excitation. LVG models appear to describe the CO
excitation in high-redshift galaxies on global scales fairly well.

As examples, the CO excitation ladders and LVG models for the
Cloverleaf quasar ($z$=2.56) and APM\,08279+5255 ($z$=3.91) are shown
in Fig.~\ref{f5}a and \ref{f5}c. Like all high-$z$ quasars studied in
detail so far, the interstellar medium in the host galaxies of these
systems is dominated by high-excitation gas components, suggesting
that the molecular gas is comparatively dense and warm, with
characteristic gas kinetic temperatures of $T_{\rm kin}$=50\,K and
densities of typically few times 10$^4$\,cm$^{-3}$ (e.g., Riechers et
al.\ 2006a, 2011b; Ao et al.\ 2008). APM\,08279+5255 is an outlier
among high-$z$ quasars, with evidence for a warm, dense gas component
with $T_{\rm kin}$=220\,K (Wei\ss\ et al.\ 2007). Typically, the
high-excitation gas components in quasars can account for all of the
flux measured in the CO($J$=1$\to$0) line as well, showing little
evidence for the existence of significant cold, low-excitation
components from the CO excitation ladders alone.  For comparison, SFGs
appear to be dominated by colder, less dense gas components with
characteristic $T_{\rm kin}$=25\,K and densities of typically few
times 10$^3$\,cm$^{-3}$ (Dannerbauer et al.\ 2009; Aravena et
al.\ 2010). Given the integrated star formation rates of SFGs, the
presence of some higher excitation gas is expected, but the relative
strength of such components is not constrained well at present. The CO
excitation in SMGs is typically intermediate between SFGs and quasars,
containing a mix of dense, high excitation gas (even though somewhat
lower excitation than in quasars on average) and spatially extended,
low excitation gas (e.g., Riechers et al.\ 2011c). The high-excitation
gas components are responsible for the dominant fraction of the line
flux in the $J$$>$2 transitions, but the low-excitation components
likely constitute a significant, and sometimes dominant fraction of
the total gas mass (Riechers et al.\ 2011c; Ivison et al.\ 2011). The
emerging picture is that the high-excitation gas components are
associated with the regions that are actively forming stars, while the
low-excitation gas is found outside these regions and represents cold,
commonly diffuse and spatially more extended reservoirs for future
star formation.

Active Galactic Nuclei (AGN) may contribute to the excitation of CO
through their intense radiation fields as well (e.g., Spaans \&
Meijerink 2008). Studies of $J$$>$10 CO lines in nearby quasars show
evidence for a likely AGN contribution to the excitation of very
high-$J$ CO lines, but the contribution to lower-$J$ CO lines is
typically minor (e.g., van der Werf et al.\ 2010). The only high-$z$
system with detected $J$$>$10 CO lines currently is APM\,08279+5255
(Wei\ss\ et al.\ 2007), and its CO excitation is not representative of
other high-$z$ quasars or SMGs with AGN in them. As such, the role of
high-$z$ AGN for the excitation of molecular gas in their hosts
remains subject to further study. ALMA will be the ideal instrument
for such investigations.

\subsection{HCN and HCO$^+$}

CO is the most commonly employed tracer of molecular gas in galaxies,
and a good tracer for the full amount of molecular gas that is
present, but it is not a particular tracer of the dense gas that is
found in actively star-forming regions. The most common tracers of
dense molecular gas in galaxies are HCN and HCO$^+$, as the critical
densities of their low-$J$ transitions\footnote{The critical densities
  of high-$J$ CO lines can approach those of HCN and
  HCO$^+$($J$=1$\to$0), but their excitation temperatures are by at
  least 1--2 orders of magnitude higher. Thus, they do not necessarily
  trace the same gas phase.} are significantly higher than those of CO
(of order 10$^5$\,cm$^{-3}$, as compared to
$\lesssim$10$^3$\,cm$^{-3}$ for CO), and more similar to the densities
found in star-forming cores. As such, the study of HCN and HCO$^+$
excitation in combination with that of CO can provide valuable
constraints on the excitation mechanisms, and it can also help to
reduce the partial degeneracies between model parameters. The only
high-redshift galaxies in which the excitation of the dense gas has
been studied to date are the Cloverleaf and APM\,08279+5255
(Fig.~\ref{f5}b and \ref{f5}d; Riechers et al.\ 2006b, 2010c, 2011a;
Wei\ss\ et al.\ 2007). The HCO$^+$ excitation in the Cloverleaf
suggest that CO and HCO$^+$ trace the same warm, dense gas phase, with
HCO$^+$ tracing the densest $\sim$15\%--20\% of the gas (Riechers et
al.\ 2011a). The HCO$^+$ excitation is comparable to what is found in
the nuclei of nearby starburst and ultra-luminous infrared
galaxies. Interestingly, the HCN excitation in APM\,08279+5255 is
poorly represented by models assuming purely collisional excitation,
but instead suggests significant enhancement by the intense, warm
infrared radiation field ($T_{\rm IR}$=220\,K) in this galaxy
(Riechers et al.\ 2010c). This is also consistent with the detection
of bright emission from multiple submillimeter H$_2$O lines, which
have critical densities of $>$10$^8$\,cm$^{-3}$, and thus are unlikely
to be collisionally excited (Fig.~\ref{f6}; Lis et al.\ 2011; Bradford
et al.\ 2011; van der Werf et al.\ 2011). These case studies
exemplify the importance of studies of dense gas excitation to
understand the conditions for star formation in high-redshift galaxies
better than possible based on the study of CO alone.

\begin{figure}[t]
\centering
\includegraphics[scale=0.22]{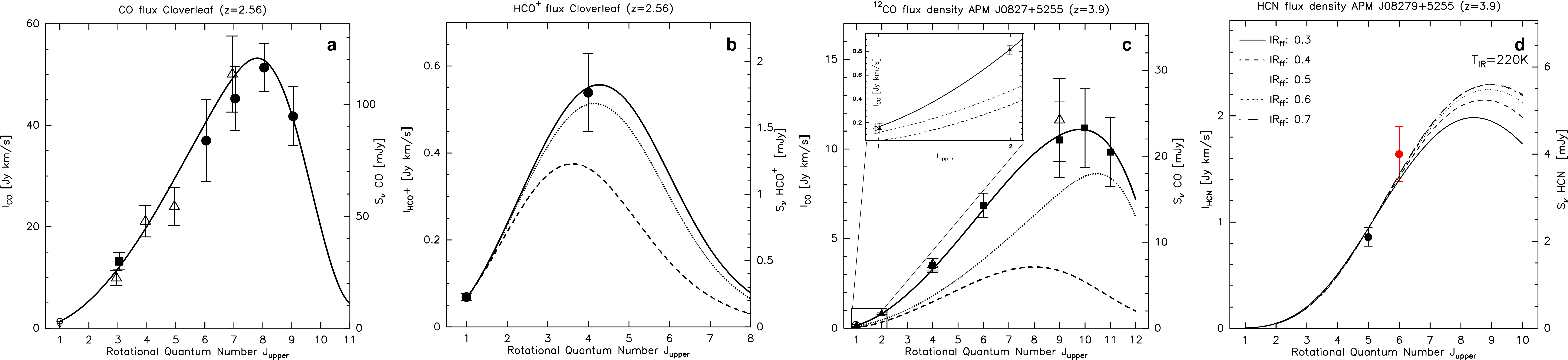} 
\caption{CO, HCO$^+$, and HCN excitation ladders (points) and LVG
  models (lines) of the molecular gas excitation for the Cloverleaf
  quasar ($z$=2.56; panels {\bf a} and {\bf b}) and APM\,08279+5255
  ($z$=3.91; panels {\bf c} and {\bf d}; Riechers et al.\ 2006b,
  2010c, 2011a; Wei\ss\ et al.\ 2007). Data for the Cloverleaf are fit
  well by a gas component with a kinetic temperature of $T_{\rm
    kin}$=50\,K and gas densities of $\rho_{\rm
    gas}$=10$^{4.5}$\,cm$^{-3}$ for CO and a 2$\times$ higher
  $\rho_{\rm gas}$ for HCO$^+$ (solid lines in panels {\bf a} and {\bf
    b}. For comparison, the dashed and dotted lines in panel {\bf b}
  show models with the same $\rho_{\rm gas}$ as in the CO model
  (dotted line:\ assuming a 10$^{0.5}$$\times$ higher HCO$^+$
  abundance). CO data for APM\,08279+5255 are fit well by two gas
  components with $T_{\rm kin}$=65 and 220\,K and $\rho_{\rm
    gas}$=10$^{5.0}$ and 10$^{4.0}$\,cm$^{-3}$. HCN data are not fit
  well by models of collisional excitation, but by radiative
  excitation through an infrared radiation field with $T_{\rm
    IR}$=220\,K. Given the similar critical densities of HCN and
  HCO$^+$, these models thus reveal a clear difference between the
  excitation conditions in both galaxies.}
\label{f5}
\end{figure}

\begin{figure}
\centering
\includegraphics[scale=0.28]{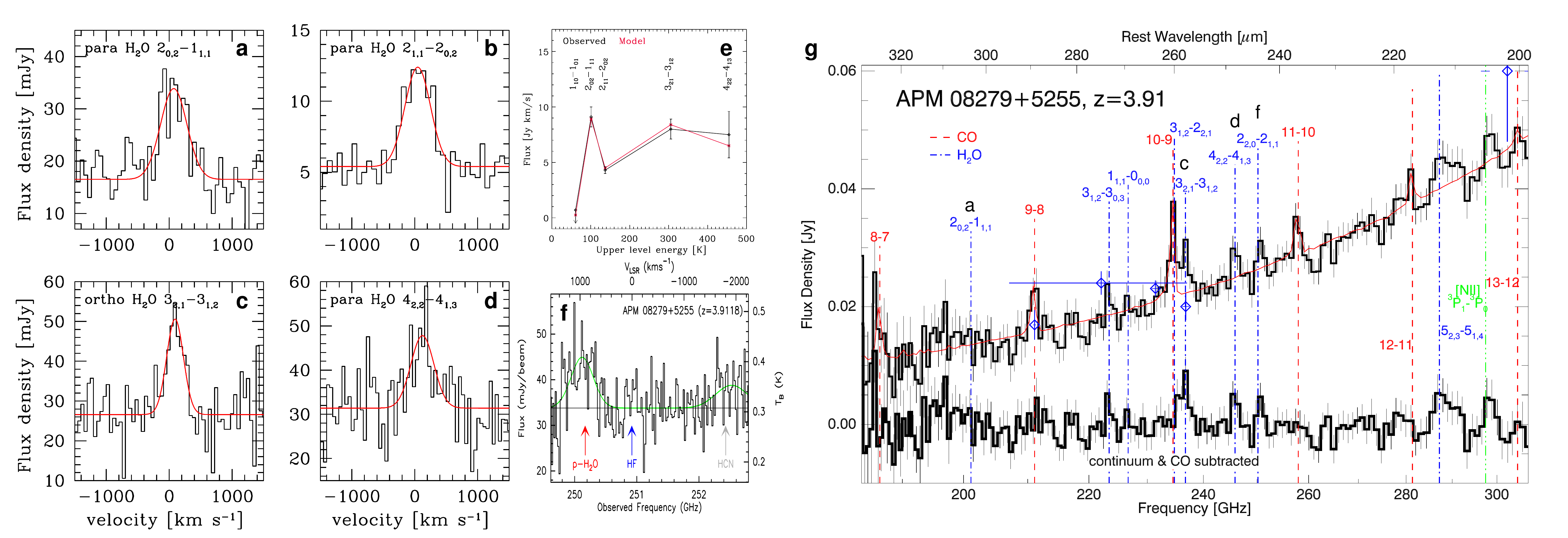} 
\caption{Line spectroscopy (panels {\bf a}--{\bf d}, {\bf f}, and {\bf
    g}) and excitation models (panel {\bf e}) of H$_2$O in
  APM\,08279+5255 ($z$=3.91; Lis et al.\ 2011; Bradford et al.\ 2011;
  van der Werf et al.\ 2011). The spectra in panels {\bf a}, {\bf c},
  {\bf d}, and {\bf f} are independent, higher spectral resolution
  interferometric observations of the corresponding lines shown in the
  broad-band spectrum in panel {\bf g}. The high H$_2$O line fluxes,
  approaching those of CO lines with three orders of magnitude lower
  critical densities, can be explained with radiative pumping of the
  H$_2$O lines by the underlying local infrared radiation field,
  rather than collisional excitation (panel {\bf e}). The inferred
  properties of the radiation field are consistent with those found to
  describe the HCN excitation in this galaxy as well (Fig.~\ref{f5};
  Riechers et al.\ 2010c).}
\label{f6}
\end{figure}

\section{Relation between Dense Molecular Gas and the Star Formation Rate}
\label{hcnlaw}

Several recent studies have concentrated on attempting to develop an
understanding for the appearance of the classical ``star formation
law'', i.e., the relation between the surface densities of the gas and
star formation rate in galaxies (e.g., Daddi et al.\ 2010a; Genzel et
al.\ 2010). Given the limited availability of high-resolution studies
to directly constrain gas and star formation rate surface densities in
high-redshift galaxies, the spatially-integrated version of this
relation still lies at the basis of much of the interpretation beyond
the local universe. For infrared-luminous galaxies, the total gas mass
and star formation rate that enter the integrated relation are most
commonly determined based on the CO line luminosity ($L'_{\rm CO}$)
and the far-infrared luminosity ($L_{\rm FIR}$; e.g., Greve et
al.\ 2005; Riechers et al.\ 2006a).

\begin{figure}[t]
\vspace{-5mm}
\centering
\includegraphics[scale=0.42]{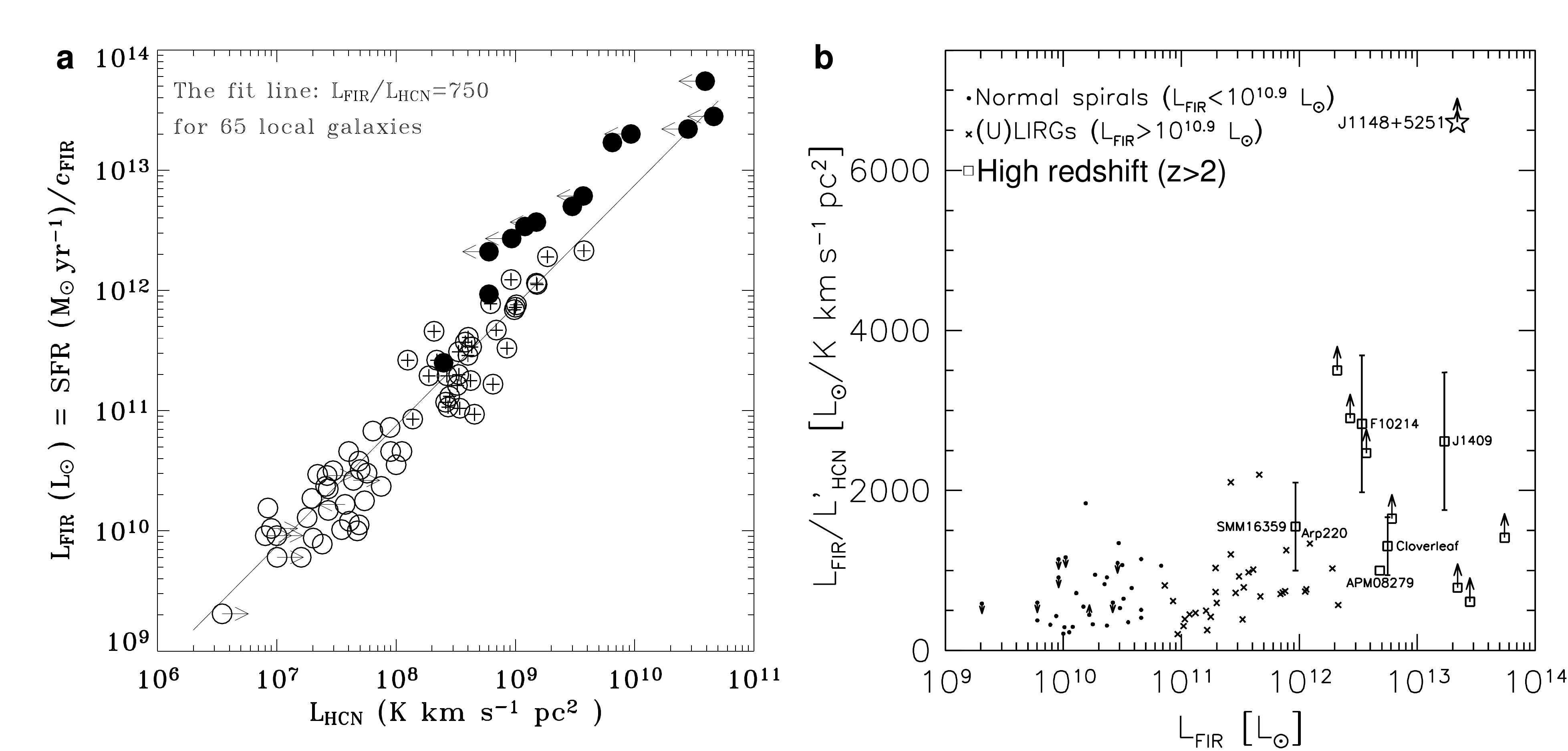} 
\caption{Relation between dense molecular gas as traced by the HCN
  line luminosity and the star formation rate as traced by the
  far-infrared luminosity. Nearby galaxies (empty and crossed circles
  in panel {\bf a}) show a linear relation between $L'_{\rm HCN}$ and
  $L_{\rm FIR}$ over many orders of magnitude, but high-redshift
  galaxies (filled circles) appear to deviate from this relation (Gao
  et al.\ 2007). Some of the most far-infrared-luminous high-redshift
  galaxies lie above the scatter in $L_{\rm FIR}$/$L'_{\rm HCN}$ of
  the most luminous nearby starbursts (panel {\bf b}; Riechers et
  al.\ 2007b).  }
\label{f7}
\end{figure}

Besides the $L'_{\rm CO}$--$L_{\rm FIR}$ relation, there is another
key relation based on spatially integrated properties.  In nearby
galaxies, the relation between the HCN luminosity ($L'_{\rm HCN}$) and
$L_{\rm FIR}$ was found to be linear (Fig.~\ref{f7}; e.g., Gao \&
Solomon 2004a, 2004b). This linear relation holds over 7--8 orders of
magnitude, ranging from Galactic dense cores to ultra-luminous
infrared galaxies (Wu et al.\ 2005). A similar relation has been
observed between the HCO$^+$ luminosity and $L_{\rm FIR}$ as well
(Riechers et al.\ 2006b). The relation between the CO and far-infrared
luminosities is much steeper, following a power-law slope of typically
1.4 when averaging over different galaxy populations (e.g., Riechers
et al.\ 2006a). This may be due to an averaging over galaxies with
both long-lasting star formation (as typically found in disk galaxies)
and merger-driven starbursts, which perhaps represent a different mode
of star formation (e.g., Daddi et al.\ 2010a, Genzel et al.\ 2010).
Galaxies with both modes of star formation can have high gas
fractions, but a different fraction of the gas itself may be actively
involved in star formation. The latter can be constrained based on
spatially resolved observations of multiple CO transition lines, which
allow to differentiate between relatively cold and diffuse gas and the
denser gas more directly related to ongoing star formation (e.g.,
Riechers et al.\ 2011c; Ivison et al.\ 2011). Dense gas tracers like
HCN (and HCO$^+$) offer an alternative, independent method to
understand this difference in lieu of detailed, spatially resolved
information.

This method is based on the linear $L'_{\rm HCN}$--$L_{\rm FIR}$
relation. The origin of this relation is thought to be that HCN traces
only the dense fraction of the gas that is associated with the
star-forming regions, which also exhibit the far-infrared emission,
but not the more diffuse cold gas not directly associated with these
regions. As such, a combination of CO (tracing all gas) and HCN
(tracing only the dense gas in star-forming regions) can provide a
more comprehensive understanding of the nature of the ``star formation
law''. Consistent with this picture, nearby ``normal'' spiral galaxies
show low $L'_{\rm HCN}$/$L'_{\rm CO}$ of typically only $\sim$4\%,
while the most luminous nearby starburst show ratios of typically
$\sim$17\% (e.g., Gao \& Solomon 2004b, Riechers et
al.\ 2007b). $L'_{\rm HCN}$/$L'_{\rm CO}$ thus may be considered a
tracer of the dense gas fraction, and its increase towards galaxies
with higher star formation rates can explain the elevated $L_{\rm
  FIR}$/$L'_{\rm CO}$ in starbursts, as well as the overall
non-linearity of the $L'_{\rm CO}$--$L_{\rm FIR}$ relation. As a
result, there is no clear indication for separate trends for spirals
and starbursts in the $L'_{\rm HCN}$--$L_{\rm FIR}$ relation, and the
overall relation appears linear. Thus, we expect that high-$z$ disk
galaxies show lower $L'_{\rm HCN}$/$L'_{\rm CO}$ ratios than nearby
starbursts with comparable $L_{\rm FIR}$ and/or star formation
rates. Such measurements will become possible with ALMA in the near
future.

Interestingly, the handful of high-redshift galaxies that were
detected in HCN emission, along with a number of upper limits,
indicate that the most luminous high-redshift star-forming galaxies
show a higher $L_{\rm FIR}$ per unit $L'_{\rm HCN}$ than expected
based on the linear relation for nearby galaxies (Fig.~\ref{f7}; Gao
et al.\ 2007; Riechers et al.\ 2007b). These galaxies are all
starbursts, and the majority of them are quasar hosts. This rising
trend can be understood if the median gas density in the luminous
high-redshift galaxies is higher than in the nearby systems, and
approaches or exceeds the critical density of the HCN($J$=1$\to$0)
transition. In this case, HCN no longer traces just the dense,
star-forming cores (which have similar $L_{\rm FIR}$/$L'_{\rm HCN}$
ratios), but rather traces the bulk of the molecular gas, in a similar
fashion as CO does in galaxies with lower, more typical gas densities
(Krumholz \& Thompson 2007). This picture is supported by the finding
that $L'_{\rm HCN}$/$L'_{\rm CO}$, on average, appears to increase
with $L_{\rm FIR}$ for nearby galaxies, but does not further increase
between nearby ultra-luminous infrared galaxies and the more
far-infrared-luminous high-redshift systems (Riechers et
al.\ 2007b). More detailed studies of dense gas excitation in distant
galaxies and an extension to multiple dense gas tracers, as possible
with the JVLA and ALMA, will be crucial to further investigate this
issue.

\section{Outlook}
\label{outlook}

Due to the completion of powerful new observatories such as the JVLA
and ALMA, progress in the field of high-redshift molecular line
studies in the coming years will be multifold. Beyond a rapid increase
in the number of galaxies detected in CO emission (e.g., Riechers et
al., in prep.; Tacconi et al.\ 2013; Wei\ss\ et al.\ 2013; see also
Vieira 2013, this volume), in-depth studies at high spatial and
spectral resolution (to probe the gas distribution, morphology and
dynamics) and in multiple molecular tracers (to probe the gas
composition and excitation) will be key to develop a deeper physical
understanding of the gas properties of distant galaxies, and of their
relation to galaxies in the present-day universe (e.g., Riechers et
al.\ 2013). Another key aspect will be to study the gas content of
distant galaxies not exclusively based on CO observations of
individual, specifically selected samples, but also in a more unbiased
fashion. This will provide direct constraints on the CO luminosity
function of galaxies at high redshift, and thus, the ``cold gas
history of the universe'' (i.e., the gas mass per unit cosmic volume
as a function of redshift). Such investigations require observations
of molecular deep fields, i.e., ``blind'' CO redshift scans over
substantial cosmic volumes. Using the full capabilities of the
recently completed JVLA correlator, the first such studies are
currently underway, and will provide unique new constraints on the
evolution of galaxies through cosmic times.

\acknowledgements I would like to thank the organizers for the kind
invitation to this workshop in honor of Paul Vanden Bout, and Paul for
his countless contributions to this exciting area of research, which
have influenced my path in the field both directly and indirectly on
many occasions -- and they will certainly continue to do so. I would
also like to thank my collaborators on studies related to this
subject, in particular Manuel Aravena, Frank Bertoldi, Chris Carilli,
Pierre Cox, Emanuele Daddi, Roberto Neri, Jeff Wagg, Fabian Walter,
Ran Wang, and Axel Wei\ss.\\[-6mm]

\end{document}